\documentclass[%
preprint,
 amsmath,amssymb,
 aps,
pra,
]{revtex4-1}
\usepackage{graphics}
\usepackage{epsfig}
\usepackage{epstopdf}
\usepackage{amsmath}
\usepackage{array}
\usepackage{graphicx}
\usepackage{dcolumn}
\usepackage{bm}

\begin{document}

\preprint{APS/123-QED}

\title{ Entropy Maximization in the Emergent Gravity Paradigm}

\author{P. B. Krishna}
\email{krishnapb@cusat.ac.in}
\author{Titus K. Mathew}%
 \email{titus@cusat.ac.in}
\affiliation{%
Department of Physics, Cochin University of Science and
Technology, Kochi, India.}%


\date{\today}

\begin{abstract}
The accelerated expansion of the universe can be interpreted as a quest for satisfying holographic equipartition. It can be expressed by a
simple
law, $\Delta V = \Delta t\left(N_{surf}- N_{bulk}\right)$ which leads to the standard Friedmann equation. This novel idea suggested by Padmanabhan in the
context of general relativity has been generalized by Cai and Yang et al. to Gauss-Bonnet and Lovelock gravities for a spatially flat universe
in different methods. We investigate the consistency of these generalizations with the constraints imposed by the
maximum entropy principle. Interestingly, both these generalizations imply entropy maximization even if their basic assumptions are different.
Further, we analyze the
consistency of Verlinde's emergent gravity with the maximum entropy principle in the cosmological context. In particular, we consider the generalization
suggested by Shu and Gong, in which an energy flux through the horizon is assumed, in addition. Even though the conceptual formulations are different, these two emergent perspectives of gravity describes a
universe which behaves as an ordinary macroscopic system. 
Our results provide further support to the emergent gravity paradigm.
\begin{description}
\item[PACS numbers]
98.80.-k, 95.36.+x
\end{description}
\end{abstract}

\maketitle


\section{Introduction}
\label{intro}The deep connection between gravitational dynamics and thermodynamics motivates the emergent interpretation of gravity. Such a connection was realized
after the
discovery of black hole thermodynamics by Bekenstein and Hawking \cite{Bekenstein1,Bekenstein2,Hawking1,Hawking2}.
A great step in this field was put forward by
 Jacobson. He obtained Einstein's field equations from the fundamental Clausius relation on a local Rindler causal horizon \cite{Jacob1}. Following this,
 various schemes for relating gravity and thermodynamics were discussed for a variety of gravity theories \cite{Eling1,Paddy1}. Later, in \cite{Paddy2}
 Padmanabhan showed that Newton's law of gravity can be derived by combining the equipartition law of energy for the horizon degrees of freedom and thermodynamic relation $S=E/2T$,
where $S$ and $T$ are the entropy and temperature of the horizon and $E$ is the active gravitational mass .

 Verlinde introduced gravity as an entropic force caused due to the changes in entropy associated with the positions of material bodies \cite{verlinde}.
He derived Newton's law of gravitation and Einstein's field equations using the holographic principle and equipartition law of energy.
Using this idea, the authors of \cite{CaiE,Shu} derived Friedmann equations through different methods. In \cite{Miao}, Miao  and  Wang discussed
 the
implications of holographic dark energy in the entropic force frame work. In a recent work, Verlinde explains the possibility of a common origin for
dark matter and dark energy \cite{Verlinde1}. This entropic force formalism also attracted a lot of investigations \cite{Gao,HMoradpour,HMoradpour1,Asheykhi1,Asheykhi2,Mann}.

Most of these studies treat the gravitational field as an emergent phenomenon assuming the spacetime background as pre-existing. Recently, a more elegant
way to view gravity as an emergent phenomenon was suggested by Padmanabhan where spacetime itself is considered as an emergent structure \cite{Paddy3}.
However it is
conceptually difficult to think of time as being emerged from some pre-geometric variables. Also it is hard to imagine the space around finite gravitating
systems as emergent. But, Padmanabhan argued that these difficulties disappear in the cosmological context, when we choose the time variable as the proper
time of the geodesic observers to whom the cosmic microwave background radiation appears homogeneous and isotropic. Thus the spatial expansion of the
universe can be described as the emergence of cosmic space with the progress of cosmic time. He successfully derived the Friedmann equation of a flat FLRW
universe in general relativity, using this new idea.

Cai generalized Padmanabhan's proposal to a higher n+1 dimensional spacetime. By properly modifying the degrees of freedom and the volume increase, he also
obtained the Friedmann equation of a flat FLRW universe in Gauss-Bonnet and Lovelock gravity \cite{Cai1}. With some modifications, this procedure was
extended by
Sheykhi to derive the dynamical equations of the universe with a spatial curvature \cite{Sheykhi}. Following this,  Ali arrived at the Friedmann equations by
considering a general form of entropy \cite{Ali}. Another generalization was suggested by Yang et.al. for the flat universe \cite{Yang} which is further
extended by Wen-Yuan Ai et.al.
for the non flat case \cite{Wang}. Instead of modifying the degrees of freedom, in \cite{Yang} and \cite{Wang} the change in Hubble volume is assumed to
be proportional to a more general
function $f(\Delta N,N_{surf})$, where $\Delta N = N_{surf}-N_{bulk}$. The authors of \cite{Eune}, extended Padmanabhan's proposal to a non flat
universe by using the appropriate invariant volume instead of the volume of the sphere in a flat space. In Ref. \cite{Chang}, Padmanabhan's conjuncture
and its modified versions were obtained from Friedmann equations. This paper also discusses the difficulties in generalizing
Padmanabhan's proposal
to a non-flat universe. In \cite{Sheykhi1}, Friedmann equations are extracted in the brane world scenarios. This emergent
paradigm has been further explored by Padmanabhan and his collaborators from a variety of perspectives \cite{sumanta1,sumanta2,sumanta3}. For recent investigations on this novel idea, see
\cite{Komatsu,Zhang,Hashemi}

As is well known, every macroscopic system evolves to a state of thermodynamic equilibrium consistent with their constraints \cite{Callen1}.
The entropy of such systems
should attain a certain maximum value in the long run. In \cite{Diego1}, it is shown that our universe with a Hubble expansion history behaves as an
ordinary macroscopic
system. In a previous work, we have proved the equivalency of holographic equipartition law and the maximum entropy principle in the context of general
relativity \cite{KT}.
In this paper we study the consistency of the generalized holographic equipartition in \cite{Cai1,Yang} with the maximum entropy principle
for a spatially flat FLRW universe. It is also our great interest to see whether Verlinde's hypothesis implies entropy maximization. We analyze
Verlinde's proposal in the cosmological context following the method in \cite{Shu}, suggested by Shu and Gong and will check its consistency with the
maximum entropy principle.

The paper is organized as follows. In the upcoming section, we obtain the constraints imposed by the generalized second law (GSL) and the maximum entropy
principle for a spatially flat FLRW universe in $n+1$ dimensional Einstein's gravity, Gauss-Bonnet gravity and Lovelock gravity. We will also prove
the consistency of an asymptotically de Sitter universe with these constraints obtained. In section 3, we analyze whether the modified
equipartition law in \cite{Cai1,Yang} ensures the maximization of entropy. In section 4, we will check the consistency of Verlinde's entropic force
formalism with the maximization entropy principle in the cosmological context. Section 5 compares
Padmanabhan's proposal with Verlinde's hypothesis and section 6 is devoted to our conclusions.

\section{Maximization of Entropy}
An ordinary, isolated macroscopic system spontaneously evolve to an equilibrium state of maximum entropy consistent with their constraints. This implies,

\begin{equation}\label{eqn:conditions1}
 \dot S \geq 0, \, \textrm{always}
\end{equation}

\begin{equation}\label{eqn:conditions2}
   \, \, \, \, \, \,  \ddot S <0 \, \, \, \textrm{at least in long run},
\end{equation}
where $'S'$ denotes the total entropy of the universe and dots indicate the derivatives with respect to a relevant variable like cosmic time. Based on
the Hubble expansion history, Pavan and Radicella have shown that our universe behaves as an ordinary macroscopic system that proceeds to a maximum entropy
state \cite{Diego1}. In this section we extend this procedure to  $n+1$ dimensional Einstein's
gravity, Gauss Bonnet gravity
and Lovelock gravity for a spatially flat universe.

The total entropy of the universe, $S$  can be
approximated as the horizon entropy since the entropy contribution
from all other components is negligibly small \cite{Egan1}. Using Bekenstein
result, the
horizon entropy can be expressed as\cite{Bekenstein1,Hawking1},

\begin{equation}
 S = \frac{A_{n+1}}{4L^{n-1}_p},
\end{equation}
where $A=n\Omega_n/{H}^{n-1}$, for $n\geq 3$, with $\Omega_n$ being the volume of the unit n-sphere. Here $L_p$ represents the Plank length and $H$ is
the Hubble parameter. The rate of
change of entropy with respect to the cosmic time is,
\begin{equation}\label{eqn:dns}
 \dot S = -\frac{n(n-1)\Omega_n }{4L^{n-1}_p} \frac { \dot H}{H^n}.
\end{equation}
Since the Hubble parameter $H$ is always positive for an expanding universe, the horizon entropy will not decrease, if $\dot H\leq0.$ The measurements
on the Hubble parameter
 \cite{Simon1,Stern1} and the numerical simulations\cite{Craw1,Carva1} have confirmed that, $\dot H <0.$
 Hence the entropy of the universe will never decrease. Even though this implies a possible thermodynamic evolution,
 whether it attains a state of equilibrium or not, is determined by the constraint on the second derivative of the entropy.
 The system approaches an equilibrium state, if the entropy corresponding to it is maximum. A maximum entropy state is
 characterized by the inequality, $\ddot S <0$ to be satisfied at least in the long run.
From equation (\ref{eqn:dns}), we have the second derivative of entropy,
\begin{equation}\label{eqn:ddns}
 \ddot S = \frac{n(n-1)\Omega_n }{4L^{n-1}_p} \left[ \left(\frac{n {\dot H}^2}{H^{n+1}}\right) - \left(\frac{\ddot H}{H^n} \right) \right].
\end{equation}
Then, one can immediately find the constraint for entropy maximization,
\begin{equation} \label{eqn:ineq2}
 n \left(\frac{\dot H^2}{H^{n+1}}\right) < \left(\frac{\ddot H}{H^n} \right),
\end{equation}
in the asymptotic limit.
As per the observational data, we have, ${\ddot H}> 0$ and $\dot H \to 0 $ in the asymptotic
limit\cite{Simon1,Stern1,Craw1,Carva1} and thus the above inequality holds true for an expanding
universe. This shows the consistency of a $n+1$ dimensional flat FLRW universe in Einstein's gravity with the
maximum entropy principle.

So far we have discussed the conditions for entropy maximization in Einstein's gravity, where the horizon entropy
follows the area law. But in Gauss-Bonnet
and Lovelock gravity, horizon entropy takes more complex form.
Hence it is worth noting, the conditions for entropy maximization in those gravity theories.

The Gauss-Bonnet gravity is a
natural extension of Einstein's gravity such that, the Gauss-Bonnet action consist of second oder terms, in addition. The entropy-area law of the
spherically
symmetric and static black hole in Gauss-Bonnet gravity is assumed to be satisfied for the FRW universe also.
Following this, the entropy
of the Hubble horizon can be expressed as\cite{Cai2,Cai3},
\begin{equation}\label{eqn:gbs}
 S=\frac{A}{4 L^{n-1}_p}\left[1+\frac{n-1}{n-3}\frac{2\tilde\alpha}{H^{-2}}\right].
\end{equation}
Here $A=n\Omega_n/{H}^{n-1}$ and $\tilde\alpha =(n-2)(n-3)\alpha$, where $\alpha$ is the Gauss Bonnet
coefficient which is positive \cite{DGB}. Then, the rate of change of entropy can be calculated from equation (\ref{eqn:gbs}) as,
\begin{equation}\label{eqn:gbdns}
 \dot S = -\frac{n(n-1)\Omega_n }{4L^{n-1}_p H^n}  (1+{2\tilde\alpha H^2})\dot H ,
\end{equation}
where $n\geq 4$. Since $\dot H \leq 0 $ , the generalized second law will be satisfied. When $t\to\infty$, $\dot H \to 0 $ and $\dot S\to 0$ indicating a state of equilibrium in the final stage.

Taking the time derivative of equation (\ref{eqn:gbdns}), we find
\begin{equation}\label{eqn:gbddns}
\begin{split}
 \ddot S = \frac{n(n-1)\Omega_n }{4L^{n-1}_p H^{n+1}}  \left[{\dot H}^{2}[{n+(2n-4)\tilde\alpha H ^{2}}] \right. - \\
 [{H}{\ddot H}(1+{2\tilde\alpha \ H^{2})}]\left. \right]
\end{split}
 \end{equation}
Since $\ddot H$ is always positive as per the observational data, the
 entropy maximization demands,
\begin{equation}\label{eqn:con3}
  |{\dot H}^{2}[{n+(2n-4)\tilde\alpha H ^{2}}]|<|{H}{\ddot H}(1+{2\tilde\alpha \ H^{2})}|
\end{equation}
in the final stage. Since ${\dot H}\to 0$, in the asymptotic limit the above inequality holds and the entropy of the apparent horizon will never
grow unbounded.

We will now move to the more general Lovelock gravity. Lovelock gravity\cite{Love1}
is a generalization of Gauss-bonnet gravity, such that the Lagrangian consists of dimensionally extended Euler densities. The entropy of the horizon in
Lovelock gravity is assumed to be of the same form as the entropy of black hole
and is given by,
\begin{equation}\label{eqn:lls}
 S= \frac{A}{4L^{n-1}_p} \sum_{i=1}^m \frac{i(n-1)}{(n-2i+1)} \hat{c_i}{H}^{2i-2}
\end{equation}
 where $\, \, \, \displaystyle m=n/2$ and the coefficients $\, \, \, \, \hat c_i$ are given by $\displaystyle \, \, \, \, \, \hat c_0=\frac{c_0}{n(n-1)} ,
 \hat c_1 =1,  \hat c_i = c_i \prod_{j=3}^m(n+1-j) \, i>1$. One can obtain the rate of change of horizon entropy in Lovelock gravity from equation
(\ref{eqn:lls}) as,
\begin{equation}\label{eqn:llds}
 \dot S= -\frac{n(n-1)\Omega_n {\dot H}}{4L^{n-1}_p H^{n+2}} \sum_{i=1}^m {i} \hat{c_i}{H}^{2i}
\end{equation}

Since ${\dot H}\leq0$ for an asymptotically de Sitter universe, the above equation ensures its consistency with the Generalized Second Law. Also when
$t\to\infty$, ${\dot H}\to0$, indicating a state of equilibrium in the final stage.
Now we will consider the second derivative of entropy by differentiating equation (\ref{eqn:llds}),
\begin{equation}\label{eqn:lldds}
\ddot S= \frac{n(n-1)\Omega_n }{4L^{n-1}_p} \sum_{i=1}^m i \hat{c_i} H ^{2i}[{(n+2-2i) \frac{{\dot H}^2}{H^{n+3}}} -\frac{\ddot H}{H^{n+2}}]
\end{equation}

Since $\ddot H$ is always positive, the horizon entropy tends to some
maximum value if,
\begin{equation}\label{eqn:con4}
  |\sum_{i=1}^m  i\hat{c_i}{H}^{2i}(n+2-2i){\frac{{\dot H}^2 }{H^{n+2}}}|<|\sum_{i=1}^m  i\hat{c_i} H^{2i}\frac{\ddot H}{H^{n+2}}|
\end{equation}
in the long run.
Since ${\dot H}\to 0$, when $t\to\infty$, the above inequality holds true in the last stage of evolution indicating the entropy maximization.

Here we have discussed the constraints imposed by the GSL and the maximum entropy principle for a flat FLRW universe in $n+1$ dimensional
Einstein, Gauss Bonnet and Lovelock
gravity. These constraints are generally satisfied by an asymptotically de Sitter universe. This entropy maximization in
the final de Sitter epoch has been already emphasized in several previous
studies \cite{SMC,Alb,Dyson} from different perspectives. It is worth mentioning that in Ref. \cite{Moradpour}, the rise of complexity content and the validity of GSL
demands an asymptotically de Sitter universe with equation of state parameter, $\omega\geq -1$. On the other hand, in \cite{Radicella},\cite{Moradpour1}, the authors
discuss the thermodynamic motivation for the existence of dark energy. The dark energy component also seems unavoidable for the attainment of equilibrium
in the brane world scenario \cite{Radicella1}. Since the attainment of equilibrium demands an asymptotically de Sitter universe, our results also point in the same
direction.
\section{Holographic equipartition and Entropy Maximization}
Our aim here is to see whether the generalized holographic equipartition in \cite{Cai1} and \cite{Yang} leads to the maximization of entropy. In \cite{KT}, it is shown that the holographic
equipartition law suggested by Padmanabhan for a flat universe effectively implies entropy maximization in the context of Einstein's gravity.
 In this section we extend this procedure for a spatially flat FLRW universe in $n+1$ Einstein's gravity, Gauss Bonnet gravity and Lovelock gravity.

\subsection{Friedmann equations from emergence of space}
Here we will  briefly review the necessary background to which our work is closely related.
We  start with Padmanabhan's idea of holographic equipartition which is followed by a brief discussion on the generalized
holographic equipartition law in \cite{Cai1,Yang}.

Padmanabhan observed that, a pure de Sitter universe with a Hubble constant H obeys holographic
principle in the form,
\begin{equation}\label{eqn:eql}
 N_{surf} = N_{bulk}
\end{equation}
Here $N_{surf}$ is the number of degrees of freedom on the Hubble horizon with radius $H^{-1}$ and is given by,
\begin{equation}\label{eqn:Nsurf1}
 N_{surf} = \frac{4 \pi}{ L^2_p H^2}.
\end{equation}
where $L^2_p$ is the Plank area, attributed to one degree of freedom. $N_{bulk}$ denotes the effective number of degrees of freedom residing in the region
enclosed by the horizon and is given by,
\begin{equation}\label{eqn:Nbulk1}
N_{bulk} =\frac{|E|}{\frac{1}{2} {k_B T }}.
\end{equation}
where $|E|=|\rho+3p|V$, the Komar energy contained inside the Hubble volume, $V=\frac{4\pi}{ 3 H^3}$; $k_B$, the Boltzmann constant and $T = \frac{H}{2\pi},$
the Gibbon's Hawking temperature.
From the above equations, one could reach at the condition, $|E|=\frac{1}{2}{N_{surf}k_B T }$.
This equality can be called as holographic equipartition since it relates the degrees of freedom in the bulk region of space, determined by the
equipartition condition to the degrees of freedom on its boundary surface.

Even though the present universe is not exactly de Sitter, many of the astronomical observations indicate that it is proceeding to a pure de Sitter state.
Based on these facts, it is suggested that the accelerated expansion of the universe can be explained as a quest for satisfying holographic equipartition.
It can be expressed by a simple law,
\begin{equation} \label{eqn:dVdt1}
 \frac{dV}{dt} ={L^2_p(N_{surf}-  N_{bulk})}.
\end{equation}
where $V$ is again the Hubble volume and $t$ is the cosmic time in Planck units $(k_B=c= \hslash =1)$. Substituting for each term in (\ref{eqn:dVdt1}) and by integrating
with the help of continuity equation one gets the standard Friedmann equation.

We will now turn our attention to the generalized holographic equipartition law suggested by Cai \cite{Cai1}.
 For a spatially flat FLRW Universe in n+1 dimensional space time, the surface degrees of freedom can be defined as,
\begin{equation}\label{eqn:Nsurf3}
 N_{surf} =\frac{ \alpha A}{ L^{n-1}_p}
\end{equation}
where $A=n\Omega_n/{H}^{n-1}$, and
\begin{equation}\label{eqn:Nbulk3}
 N_{bulk} =\frac{-4\pi\Omega_n}{H^{n+1}} \frac{(n-2)\rho +np }{n-2}.
\end{equation}
Also the equipartition law given in (\ref{eqn:dVdt1}) is modified for an (n+1) dimensional space as,
 \begin{equation} \label{eqn:dVdt3}
 \alpha\frac{dV}{dt} ={L^{n-1}_p(N_{surf}- N_{bulk})}.
\end{equation}
where, $V=\Omega_n /{H}^{n}$, the volume of the n-sphere.
Substitution  of the degrees of freedom in equations (\ref{eqn:Nsurf3}) and (\ref{eqn:Nbulk3}) in the holographic equipartition law given in
(\ref{eqn:dVdt3}) gives  the Friedmann equation of the flat FLRW universe in n+1 dimensional space time.
Now we will briefly describe the extension of the above method to Gauss-Bonnet gravity and more general Lovelock gravity.

In Gauss-Bonnet gravity,
the effective area corresponding to the horizon entropy can be defined from equation (\ref{eqn:gbs}) as,
\begin{equation}
 \tilde A=A\left[1+\frac{n-1}{n-3}\frac{2\tilde\alpha}{H^{-2}}\right].
\end{equation}
 The  corresponding increase in effective volume is given by,
\begin{equation}\label{eqn:Veff1}
 \frac{d\tilde V}{dt} =-\frac{n\Omega_n}{H^{n+1}}(1+2\tilde\alpha H^{2})\dot H
\end{equation}
From the above expression, the degrees of freedom at the apparent horizon can be assumed as,
\begin{equation}\label{eqn:Nsurf4}
  N_{surf}=\alpha \frac{ n\Omega_n }{{H^{n+1}}{L^{n-1}_p}}(H^{2}+\tilde\alpha H^{4}).
\end{equation}
The degrees of freedom in the bulk is still given by (\ref{eqn:Nbulk3}). Then, from
equation (\ref{eqn:dVdt3})  the Friedmann equation of the flat FLRW Universe in Gauss-Bonnet gravity can be derived.

In Lovelock gravity, the increase in effective volume can be calculated from (\ref{eqn:lls}) as,
\begin{equation}\label{eqn:Veff2}
 \frac{d\tilde V}{dt}= -\frac{n\Omega_n}{H^{n+3}}
 \left(\sum_{i=1}^m i \hat{c_i}{H}^{2i}\right)\dot H
\end{equation}
From the above equation the surface degrees of freedom can be defined as,
\begin{equation}\label{eqn:Nsurf5}
 N_{surf}= \alpha \frac{ n\Omega_n }{H^{n+1}{L^{n-1}_p}}  \sum_{i=1}^m  \hat{c_i}{H}^{2i}
\end{equation}
As described earlier from (\ref{eqn:Nbulk3}), (\ref{eqn:Veff2}), (\ref{eqn:Nsurf5}) and (\ref{eqn:dVdt3}) one  can arrive at the Friedmann equation of a
flat FLRW universe in Love1ock gravity.

Inspired by this work, Yang et al. proposed another generalization of Padmanabhan's holographic equipartition law \cite{Yang}. In this case, the surface degrees of
freedom on the Hubble surface is assumed to be proportional to the area of the surface regardless of the gravity theory while the bulk degrees of
freedom obeys the equipartition law of energy. Here the emergence of cosmic space is described by a general form of a dynamical equation,
\begin{equation}\label{eqn:yang1}
 \frac{dV}{dt} = L^{n-1}_p f(\Delta N,N_{surf}),
\end{equation}
where, $V=\Omega_n /{H}^{n}$, irrespective of the gravity theory and $\Delta N = N_{surf}-N_{bulk}$.
The authors have used the relations in (\ref{eqn:Nsurf3}) and (\ref{eqn:Nbulk3}) for defining $N_{surf}$ and $N_{bulk}$ respectively in an $n+1$
dimensional space time. If $f(\Delta N,N_{surf})$ is chosen to be the most simplest form
 $f(\Delta N) = \frac{\Delta N}{\alpha}$, as in \cite{Paddy2,Cai1},
one could arrive at the Friedmann equations for a flat universe in Einstein's gravity. In order to derive the Friedmann
 equation in Gauss-Bonnet gravity one has to choose,
\begin{equation}\label{eqn:yang2}
 f(\Delta N,N_{surf})= \frac{\Delta N /\alpha+{\tilde \alpha K(N_{surf}/\alpha})^{1+\frac{2}{1-n}}}{1+2 \tilde \alpha K (N_{surf}/\alpha)^{\frac{2}{1-n}}},
\end{equation}
where $K = (n\Omega_n/L^{n-1}_p)^{\frac{2}{n-1}}$ and $\tilde \alpha$ is a parameter with length dimension 2. If $f(\Delta N,N_{surf})$ is assumed to be a
more general function,
\begin{equation}\label{eqn:yang3}
 f(\Delta N,N_{surf})= \frac{\Delta N /\alpha+{\sum_{i=2}^m\tilde {c_i} K_i(N_{surf}/\alpha})^{1+\frac{2i-2}{1-n}}}{1+\sum_{i=2}^m i \tilde {c_i} K_i (N_{surf}/\alpha)^{\frac{2i-2}{1-n}}},
\end{equation}
where $K_i = (n\Omega_n/L^{n-1}_p)^{\frac{2i-2}{n-1}}$, $m=[n/2]$ and $\tilde c_i$ are some coefficients with the
length dimension $(2i-2)$; one can
arrive at the Friedmann equation of the $n+1$ dimensional spatially flat FLRW universe in Lovelock gravity.

\subsection{Holographic equipartition and Entropy maximization: Analysis of Cai's proposal}
Let us consider a spatially flat $n+1$ dimensional universe in Einstein's gravity. The time derivative of the cosmic volume,
 $V=\Omega_n /{H}^{n}$, can be calculated as,
\begin{equation}
  {dV\over dt}= -n\Omega_n \frac{\dot H}{H^{n+1}}
\end{equation}
Now recalling equation (\ref{eqn:dns}), we can write
\begin{equation}\label{volume change in terms entropy1}
 {dV\over dt} = \frac{4L^{n-1}_p }{H(n-1)}  \dot S
\end{equation}
Then, holographic equipartition law in equation (\ref{eqn:dVdt3}), can be written as,
\begin{equation}\label{Eeqncon1}
 \dot S ={\frac{(n-2)H}{2}(N_{surf}- N_{bulk})}.
\end{equation}
For an expanding universe, we have ${dV\over dt}\geq0$, which demands
\begin{equation}
{N_{surf}- N_{bulk}}\geq0
\end{equation}
This in turn ensures the non-negativity of the r.h.s. of equation (\ref{Eeqncon1}). Consequently the generalized second law in the form $\dot S\geq0$ will be
satisfied.

Taking the differential of (\ref{Eeqncon1}), we get
\begin{equation}\label{Eeqncon11}
\begin{split}
 \ddot S ={\frac{(n-2)\dot H}{2}(N_{surf}- N_{bulk})}+{{\frac{(n-2)H}{2}}} \\ {{d\over dt}(N_{surf}- N_{bulk})}
\end{split}
 \end{equation}
We have in the asymptotic limit, $N_{bulk}\to N_{surf}$ and the first term in the above expression vanishes. As per the holographic equipartition law, the
evolution of the universe can be explained as a tendency to equalize the degrees of freedom on the horizon and that in the bulk. In other words, one can say
that the universe is trying to minimize the holographic discrepancy with the progress of time. Since $ {N_{bulk}}$ can not exceed $N_{surf}$, we have
$'{N_{surf}- N_{bulk}}'$, always positive and tending to zero in the asymptotic limit. This in turn implies that
\begin{equation}
 {d\over dt}(N_{surf}- N_{bulk})<0
\end{equation}
 which guarantees the non-positivity of $\ddot S$ in the long run, ensuring the consistency with the maximum entropy principle.
 Also, substituting (\ref{eqn:Nsurf3}) and (\ref{eqn:Nbulk3}) in (\ref{Eeqncon11}) we get the condition for  $\ddot S<0$ as,
\begin{equation}
 n \left(\frac{\dot H^2}{H^{n+1}}\right) < \left(\frac{\ddot H}{H^n} \right),
\end{equation}
This is nothing but the constraint in equation (\ref{eqn:ineq2}) we  obtained in the last section for the entropy maximization which is satisfied by an asymptotically de Sitter
universe.

Next, we will investigate whether the holographic equipartition law proposed in the context of Gauss-Bonnet gravity implies entropy maximization. Combining
(\ref{eqn:Veff1}) and (\ref{eqn:gbdns}), one can relate the rate of change of effective volume within the apparent horizon to the rate of change of entropy
as,
\begin{equation}
 {d\tilde V\over dt} = \frac{4L^{n-1}_p }{H (n-1)} \dot S.
\end{equation}
Note that the above equation is in the same form of equation (\ref{volume change in terms entropy1}) with $V$ and $\dot S$ are replaced with the
corresponding expressions in Gauss-Bonnet gravity.
Hence, the holographic equipartition law in equation (\ref{eqn:dVdt3}) can be rewritten as,
\begin{equation}\label{gbeqncon1}
 \dot S ={\frac{(n-2)H}{2}(N_{surf}- N_{bulk})}.
\end{equation}
just as in the case before, where ${N_{surf}}$ and $ {N_{bulk}}$ are given by equation (\ref{eqn:Nsurf4}) and equation (\ref{eqn:Nbulk3}) respectively.
For an expanding universe, we have ${dV\over dt}\geq0$, which demands ${N_{surf}- N_{bulk}}\geq0$. Hence the above equation guarantees the consistency
of the universe with the generalized second law.  Now we have the second derivative of entropy,
\begin{equation}\label{GBeqncon11}
\begin{split}
 \ddot S ={\frac{(n-2)\dot H}{2}(N_{surf}- N_{bulk})}+{{\frac{(n-2)H}{2}}} \\ {{d\over dt}(N_{surf}- N_{bulk})}
 \end{split}
\end{equation}
as we have seen earlier. Here also, the first term vanishes in the asymptotic limit and the entropy tends to a certain maximum value if,
$ {d\over dt}(N_{surf}- N_{bulk})<0$

Since, the universe is trying to decrease the holographic discrepancy with the progress of time, the above inequality holds and the entropy gets saturated
in the asymptotic limit. Substituting the equations (\ref{eqn:Nsurf4}) and  (\ref{eqn:Nbulk3}) in (\ref{GBeqncon11}), we get the constraint for the non positivity of $\ddot S$,
\begin{equation}
  |{\dot H}^{2}[{n+(2n-4)\tilde\alpha H ^{2}}]|<|{H}{\ddot H}(1+{2\tilde\alpha \ H^{2})}|
\end{equation}
which is same as the inequality in (\ref{eqn:con3}) that we  obtained earlier for the entropy maximization in Gauss-Bonnet gravity.

Let us now generalize this procedure to Lovelock gravity. Recalling the equations (\ref{eqn:dVdt3}),
(\ref{eqn:Veff2}) and(\ref{eqn:llds}), the holographic equipartition law
in Lovelock gravity can be expressed as,
\begin{equation}\label{LLeqncon1}
 \dot S ={\frac{(n-2)H}{2}(N_{surf}- N_{bulk})}.
\end{equation}
 where ${N_{surf}}$ and $ {N_{bulk}}$ are given by equation (\ref{eqn:Nsurf5}) and equation (\ref{eqn:Nbulk3}) respectively. Note that the above relation
 takes the same form as in the Einstein's and Gauss-Bonnet gravity. We have ${N_{surf}- N_{bulk}}\geq0$ for an expanding universe which guarantees the
 consistency with the generalized second law, $\dot S\geq 0$.
The second derivative of entropy takes the form,
\begin{equation}\label{LLeqncon111}
\begin{split}
 \ddot S ={\frac{(n-2)\dot H}{2}(N_{surf}- N_{bulk})}+{{\frac{(n-2)H}{2}}} \\ {{d\over dt}(N_{surf}- N_{bulk})}
 \end{split}
\end{equation}
 Here, $\ddot S$ will be negative in the asymptotic limit, if ${d\over dt}(N_{surf}- N_{bulk})<0 $, just as in the previous case. As the holographic discrepancy is a decreasing function of time, the above inequality will be satisfied in the long run.
Substituting (\ref{eqn:Nsurf5}) and (\ref{eqn:Nbulk3}) in equation (\ref{LLeqncon111}), we obtained the constraint for entropy maximization,
\begin{equation}
  |\sum_{i=1}^m  i\hat{c_i}{H}^{2i}(n+2-2i){\frac{{\dot H}^2 }{H^{n+2}}}|<|\sum_{i=1}^m  i\hat{c_i} H^{2i}\frac{\ddot H}{H^{n+2}}|
\end{equation}
As expected, this is same as the constraint (\ref{eqn:con4}) we have obtained for the entropy maximization in Lovelock gravity in the previous section.
Thus, the validity of the holographic equipartition law ensures the validity of maximum entropy principle. The tendency for satisfying the holographic
equipartition
can be explained as a tendency for maximizing entropy in Einstein, Gauss Bonnet and Lovelock gravity theories for a flat FLRW universe. Since the law of
emergence has the same form in Cai's proposal irrespective of the gravity theories, the equations (\ref{Eeqncon1}),(\ref{gbeqncon1}),(\ref{LLeqncon1})
and their derivatives take the same form, although the definitions of each term in those equations are different.
The above discussions strengthens the deep connection between the emergence of space and the entropy maximization.

\subsection{Holographic equipartition and Entropy maximization: Analysis of Yang et.al.'s proposal}
Even though Cai obtained Friedmann equations for a flat universe in Gauss-Bonnet and Lovelock gravity, his work was criticized for using effective volume
for the volume change and plain ordinary volume for defining the bulk degrees of freedom. In order to overcome this discrepancy
 Yang et.al. used the plain ordinary volume for defining both the rate of emergence and the bulk degrees of freedom \cite{Yang}.
 In the context of general relativity, the generalized holographic law
given in equation (\ref{eqn:yang1}) is not different from Cai's proposal. Then, as we have seen earlier the holographic discrepancy, $\Delta N$ vanishes in the
long run ensuring the consistency with the generalized second law and the maximum entropy principle. On the other hand, in Gauss-Bonnet and Lovelock
gravity,'$\Delta N$' is generally non-vanishing. Even in the final de Sitter state, the holographic discrepancy will not be zero. In this case, the
emergence of cosmic space couldn't be explained as a tendency for equalizing the degrees of freedom. Hence it is worth investigating whether this
generalization fulfills the generalized second law and the maximum entropy principle.

Now, from equations (\ref{eqn:gbdns}), (\ref{eqn:yang1}) and (\ref{eqn:yang2}), the rate of change of entropy with respect to the cosmic time can be expressed as,
\begin{equation}
 \dot S=\frac{(n-1)H}{4}(1+2\tilde\alpha H^{2}) f(\Delta N,N_{surf}).
\end{equation}
Since $dV/dt\geq 0$, for an expanding universe, equation (\ref{eqn:yang1}) guarantees the non-negativity of $f(\Delta N,N_{surf})$ and thus ensures the consistency
with the generalized second law. Differentiating the above equation, we get
\begin{equation}
\begin{split}
 \ddot S=\frac{(n-1)}{4}\frac{d}{dt}(H(1+2\tilde\alpha H^{2})) f(\Delta N,N_{surf})+ \\
 \frac{(n-1)}{4} H (1+2\tilde\alpha H^{2}) \frac{d}{dt}(f(\Delta N,N_{surf}))
\end{split}
 \end{equation}
In order to fulfill the maximum entropy principle, the r.h.s. of the above equation must be negative in the long run. In the asymptotic limit, when
$t\to\infty$, the rate of emergence of space, $dV/dt$ will tend to zero. Hence as per equation (\ref{eqn:yang1}), $f(\Delta N,N_{surf})\to 0$ and the first term
in the above expression vanishes. Since the rate of emergence is always positive, tending to zero in the long run, from equation (\ref{eqn:yang1}), we get
\begin{equation}
\frac{d}{dt}(f(\Delta N,N_{surf})<0
\end{equation}
in the final stage. This guarantees the non positivity of '$\ddot S$' in the last stage of evolution and thus ensures the consistency with the maximum
entropy principle.

In Lovelock gravity, from equations (\ref{eqn:llds}), (\ref{eqn:yang1}) and (\ref{eqn:yang3}), the change in entropy can be obtained as,
\begin{equation}
 \dot S=\frac{(n-1)}{4}\sum_{i=1}^m  i\hat{c_i} H^{2i-1} f(\Delta N,N_{surf}).
\end{equation}
and its derivative,
\begin{equation}
\begin{split}
 \ddot S=\frac{(n-1)}{4}\frac{d}{dt}(\sum_{i=1}^m  i\hat{c_i} H^{2i-1}) f(\Delta N,N_{surf})+ \\
 \frac{(n-1)}{4}\sum_{i=1}^m  i\hat{c_i} H^{2i-1}
 \frac{d}{dt} (f(\Delta N,N_{surf}))
 \end{split}
\end{equation}
As per the earlier arguments, the above equations guarantee the validity of the generalized second law and the maximum entropy principle. What is striking
in our result is that even if the law governing the emergence of space in \cite{Yang} does not lead to the condition $N_{surf}=N_{bulk}$, it guarantees
the validity of the generalized second law and the maximum entropy principle. In short, a flat FLRW universe that obeys the generalized holographic
equipartition law in \cite{Cai1}\cite{Yang}, behaves as an ordinary macroscopic system in the context of Einstein, Gauss-Bonnet and Lovelock gravity. In the light of
above discussions, the achievement of holographic equipartition could be interpreted as the attainment of maximum entropy in a spatially flat universe.
\section{Entropy Maximization in Verlinde's emergent gravity}
In \cite{verlinde}, Verlinde introduced the treatment of gravity as an emergent phenomenon. He interprets gravity as an entropic force experienced by a material body when it
approaches a holographic screen. We, here, consider some of the arguments in \cite{verlinde} and analyze the consistency of this proposal with the generalized second
law and the maximum entropy principle in the cosmological context.

In \cite{verlinde}, the number of bits of information on a holographic screen of area $A$ is assumed as,
\begin{equation}
 N=\frac{A c^3}{G\hbar}.
\end{equation}
Then, from the equipartition law, the total energy of the system can be calculated as,
\begin{equation}
 E=\frac{1}{2} N k_B T
\end{equation}
where $T$ is the temperature on the screen. The energy $E$ in the above expression is assumed to be equal to $M c^2$ where $M$ represents the mass that
would emerge on the part of space enclosed by the screen. Verlinde arrive at Newton's law of gravitation and  Einstein's field equations from these
postulates.

By generalizing this proposal to dynamic spacetimes Shu and Gong \cite{Shu} and Cai et.al. \cite{CaiE} derive Friedmann equations using different methods.
We assume the Hubble horizon as the boundary of the universe, as earlier, and hence follow the method of \cite{Shu}. Apart from the equipartition law and the
holographic principle the authors of \cite{Shu} assume an energy flux through the horizon. If there is an energy flux through the horizon, the energy
$\varepsilon$ enclosed by it will increase in course of time.  Consequently, as per the equipartition law, the temperature and the number of bits on the
screen changes. If the universe is assumed to be flat, the radius of the apparent horizon will be equal to the Hubble radius, $r_H$. Then, the change in total
energy in an infinitesimal interval of time $dt$ can be expressed as,
\begin{equation}\label{eqn:energy change}
 d\varepsilon = \frac{1}{2} N_H d T_H + \frac{1}{2} T_H dN_H
\end{equation}
where $N_H= \frac{4\pi {r_H}^2}{L^2_p}$ and
$T_H= \frac{\hbar}{2\pi r_H}$, the Hawking temperature.
Here the increase in the number of bits on the Hubble sphere,
\begin{equation}\label{eqn:dNH}
 d N_H =\frac{8\pi r_H}{L^2_p} d r_H
\end{equation}
and the change in Hawking temperature is,
\begin{equation}\label{eqn:dTH}
 d T_H=-\frac{\hbar}{2\pi {r_H}^2 } d r_H
\end{equation}
as in \cite{Shu}. This change in energy, $d\varepsilon$ will be equal to the energy flow through the horizon within a time interval $dt$ which is given by,
\begin{equation}\label{eqn:total energy change}
 -dE=d\varepsilon=4\pi {r_H}^3 (\rho+p) H dt
\end{equation}
Combining equations (\ref{eqn:energy change}), (\ref{eqn:dNH}) and (\ref{eqn:dTH}), we arrive at,
\begin{equation}\label{dedt}
 \frac{d\varepsilon}{dt} = -\frac{\dot H}{L^2_p H^2}
\end{equation}
From the definition of horizon entropy, $S=A/4$ the above relation can be expressed as,
\begin{equation}\label{Tdsdt}
 T \frac{dS}{dt} = -\frac{\dot H}{L^2_p H^2}
\end{equation}
With the help of Friedmann equation, the energy flux through the horizon can be defined from equation (\ref{eqn:total energy change}) as,
\begin{equation}
 \frac{d\varepsilon}{dt} = \frac{4\pi^2(1+\omega)}{H^2} \rho
\end{equation}
If the universe is assumed to be asymptotically de Sitter, the equation of state, $\omega \to -1$, when $t\to\infty$. In consequence, as per the above
equation $\frac{d\varepsilon}{dt}\to 0$. Also for $\omega\geq-1$, $\frac{d\varepsilon}{dt}\geq 0$. Here the universe is trying to minimize the energy
flux through the horizon with the progress of cosmic time. In other words, the evolution of the universe can be interpreted as a tendency for minimizing the
flux through the horizon. From equations (\ref{dedt}) and (\ref{Tdsdt}), the rate of change of entropy can be expressed as,
\begin{equation}
 \dot S =\frac{2\pi}{H} \frac{d\varepsilon}{dt}
\end{equation}
 Since $\frac{d\varepsilon}{dt}\geq 0$ for $\omega\geq -1$, the law of emergence is consistent with the GSL.

Now, we will check the convexity condition for the maximization of entropy. Differentiating the above equation once again with respect to time, we get
\begin{equation}
 \ddot S = \frac{8 \pi^2}{H^3}(1+\omega)\dot\rho -\frac{24 \pi^2 \dot H}{H^4}(1+\omega)\rho + \frac{8 \pi^2}{H^3} \dot\omega \rho
\end{equation}
In an asymptotically de Sitter universe, $\omega \to -1$, as $t \to\infty  $ and the first two terms in the above expression vanishes. Since $\dot\omega$ is
always negative the total entropy will never grow unbounded. Thus, Verlinde's proposal which is generalized in \cite{Shu} is in agreement with the
generalized second law and the maximum entropy principle.

\section{Padmanabhan's proposal Vs Verlinde's hypothesis}
Padmanabhan describes the evolution of the universe as a quest for decreasing the holographic discrepancy \cite{Paddy3}. But, based on the arguments in
\cite{Shu}, one can interpret the cosmic evolution as a tendency for reducing the energy flux through the horizon. Although the basic assumptions are different, 
the approaches in \cite{Paddy3} and \cite{Shu} leads to the same results in the context of cosmology. Both
these proposals assure the validity of GSL and the maximum entropy principle. Moreover, both of them describe a universe that proceeds to a pure de Sitter
state and thus demand the presence of dark energy which is not too different from the cosmological constant.  The authors of \cite{HMoradpour}, 
have pointed out a possible connection between Verlinde's and
Padmanabhan's arguments by proposing a generalized entropy.

It is worth mentioning that, the authors of \cite{Shu}, assume the conventional first law of black hole dynamics, $TdS=dE$, where $dE$ is the energy flux
through the horizon. This energy flux, $dE$ is getting reduced and finally vanishes, ensuring the consistency with the maximum entropy principle. Meanwhile,
Padmanabhan's holographic equipartition law could be expressed in the form, $TdS=dE_G +PdV$, where we have an extra term $PdV$. Here, 
$E_G=\frac{c^4}{G}(\frac{A_H}{16\pi})^{\frac{1}{2}}$, is the energy associated with the horizon of area $A_H$ and $PdV$ is the work function of the matter source.
For detailed discussion see \cite{Paddynew2,Kothawala}. When the universe evolves to the final de Sitter state with a constant Hubble parameter, both $dE_G$ and $PdV$ will
vanish resulting in the maximization of entropy. But, one could easily reach at the maximum entropy principle directly from the holographic equipartition law 
 \cite{KT}.

We wish to emphasize that the approach in \cite{Shu} is slightly different from Verlinde's original proposal, as the authors assume an energy flux through 
the horizon in addition.
Now, following Verlinde's proposal in \cite{verlinde}, the total energy can be defined as,
\begin{equation}\label{Verlinde relation}
 E=\frac{1}{2} N k_B T =M c^2
\end{equation}
where $N=\frac{A c^3}{G\hbar}$, the degrees of freedom on the holographic screen. In an FLRW universe '$M$' is usually taken as the Komar mass
$|\rho+3p|V$ \cite{CaiE,HMoradpour1}, instead of the total mass $\rho V$. Hence, assuming the Hubble horizon as the boundary of the universe the above
equation can be written as,
\begin{equation}\label{Verlinde relation1}
N =\frac{|E|}{\frac{1}{2} {k_B T }},
\end{equation}
where $|E|= \frac{4\pi\rho|1+3\omega|}{3 H^3}$, the Komar energy inside the Hubble volume.
 Assuming a thermal equilibrium between the horizon and the fluid inside the horizon, we take
$T=H/2\pi$, the Gibbons-Hawking temperature. Also, in Ref. \cite{Mimoso}, it is argued that even though the radiation can not reach thermal equilibrium with the
horizon, non-relativistic matter may and dark energy might. With the help of Friedmann equation, the equation (\ref{Verlinde relation1}) can be expressed as,
\begin{equation}\label{Verlinde relation 2}
 \frac{4\pi}{L^2_p H^2}= |1+3\omega| \frac{2\pi c^3}{G\hbar H^2}
\end{equation}
In the final de Sitter state $\omega\to -1$ and the r.h.s. of the above equation becomes equal to the l.h.s. Thus, in a pure de Sitter universe, the degrees
of freedom on the horizon can be defined as,
\begin{equation}
N_{surf} =\frac{|E|}{\frac{1}{2} {k_B T }}.
\end{equation}
This is nothing but Padmanabhan's holographic equipartition condition, since it relates the degrees of freedom on the surface to the degrees of freedom
determined by the equipartition condition. 

But, it has to be noted that equation (\ref{Verlinde relation 2}) holds true only in the final de Sitter state. Generally we have $\omega\geq -1$, throughout
the evolution for a universe that tends to a final de Sitter state. For instance, $\omega =\frac{1}{3}$ in the radiation dominated phase, $\omega=0$ for
the matter dominated phase and $\omega=-1$, for the dark energy dominated phase. Hence equation (\ref{Verlinde relation 2}) should be rewritten as the
inequality,
\begin{equation}\label{Verlinde relation 3}
 \frac{4\pi}{L^2_p H^2}\geq |1+3\omega| \frac{2\pi c^3}{G\hbar H^2}.
\end{equation}
From the definitions of $N_{surf}$ and $N_{bulk}$, the above inequality can be expressed as,
\begin{equation}\label{Padmanabhan relation}
 N_{surf}\geq N_{bulk}
\end{equation}
In the final de Sitter state $N_{bulk}$ approaches $N_{surf}$ and the holographic equipartition is achieved.
 Conversely, starting from Padmanabhan's argument $N_{surf}\geq N_{bulk}$ one can deduce the inequality,
\begin{equation}
N_{surf} \geq \frac{|E|}{\frac{1}{2} {k_B T }},
\end{equation}
which reduces to $E=\frac{1}{2} N k_B T =M c^2$, in the asymptotic limit. Hence if we assume the Hubble horizon as the boundary of the universe Verlinde's
assumption in equation (\ref{Verlinde relation}) can be obtained as a limiting case of Padmanabhan's relation in (\ref{Padmanabhan relation}). However, we
wish to highlight the fact that the approaches in \cite{verlinde} and \cite{Paddy3} are conceptually different and there exist no equivalency between 
them \cite{Paddynew1}.

\section{Conclusion}
In this paper, we investigate the consistency of the generalized holographic equipartition with the maximum entropy principle. In particular, we have
considered the generalizations in \cite{Cai1,Yang}, where the authors extended Padmanabhan's proposal to Gauss-Bonnet and Lovelock gravities for a
spatially flat universe. We have also analyzed the consistency of Verlinde's entropic force formalism with the maximum entropy principle in the cosmological
context.

In \cite{Diego1}, Pavon and Radicella have shown that our universe behaves as an ordinary macroscopic system that proceeds to a maximum entropy state.
But, their results are restricted to the $3+1$ dimensional Einstein's gravity. Hence we first extended the procedure in \cite{Diego1} to $n+1$
dimensional Einstein, Gauss-Bonnet and Lovelock gravities and obtained the constraints for the maximization of entropy. These constraints are generally satisfied
by an expanding universe that proceeds to a final de Sitter epoch.

One of our main aim was to check whether the generalized holographic equipartition in \cite{Cai1,Yang} imply entropy maximization. Following the
modified holographic equipartition suggested by Cai
in \cite{Cai1}, we have found that the condition $\dot S\geq 0$ implies ${N_{surf}- N_{bulk}}\geq0$ and $\ddot S<0$ (in the long run) leads to
$ {d\over dt}(N_{surf}- N_{bulk})<0$ in Einstein, Gauss-Bonnet and Lovelock gravity theories. We showed that these conditions are compatible with the respective constraints
that we have obtained for entropy maximization in each gravity theory. Thus, an asymptotically de Sitter universe which evolves to minimize the holographic descrepancy $N_{surf}- N_{bulk}$
proceeds to a maximum entropy state. On the other hand, following \cite{Yang}, we found that $\dot S\geq 0$ leads to $f(\Delta N,N_{surf})\to 0$ and
$\ddot S<0$ (in the long run) leads to $\frac{d}{dt}(f(\Delta N,N_{surf})<0$ in all these gravity theories. Since the rate of emergence is always positive
tending to zero in the final stage, $f(\Delta N,N_{surf})$ satisfies the above conditions, ensuring the entropy maximization. What is remarkable here
is that, even if the law of emergence does not guarantees the
condition $N_{surf}= N_{bulk}$, it ensures the consistency with the GSL and the maximum entropy principle. The above results provide a thermodynamic basis
for the law of emergence beyond Einstein's gravity.

It may be noted that Verlinde's entropic force formalism has been generalized to the cosmological context in different
methods. We have considered one of such generalizations
in \cite{Shu}, where the authors assume an energy flow, $d\varepsilon$ through the horizon. Following this approach, we found that this energy flow
through the horizon is getting reduced in course of time. In this case, the generalized second law,  $\dot S\geq 0$ implies the condition
$\frac{d\varepsilon}{dt}\geq 0$. This energy flow through the horizon will eventually stop in the final de Sitter state of maximum entropy. Hence it can be
argued that the universe is trying maximize its entropy by reducing the energy flow through the horizon.

Finally, we made a comparison between Padmanabhan's proposal and Verlinde's hypothesis. According to Padmanabhan, the evolution of the universe can be
interpreted as a tendency for decreasing the holographic discrepancy. But, following the argument in \cite{Shu}, one can interpret the cosmic evolution
as a tendency for reducing the energy flux through the horizon. However both these proposals ensures the consistency with the GSL and the maximum entropy
principle. Moreover, both of them demands an asymptotically de Sitter universe which in turn implies the presence of dark energy which is not too
different from the cosmological constant. We have already mentioned that the approach in \cite{Shu} is slightly different from Verlinde's original
proposal, as the authors assume an energy flux through the horizon in addition. It should also be noted that, one of the  basic assumption of Verlinde,
$\frac{1}{2} N k_B T =M c^2$ will hold true only in a pure de Sitter universe. This relation could in general be expressed as the inequality 
$N \geq \frac{Mc^2}{\frac{1}{2} {k_B T }}$ which has the same form of Padmanabhan's assumption $ N_{surf}\geq N_{bulk}$.

Although the conceptual formulations in \cite{Shu} and \cite{Paddy3} are different, both these emergent perspectives of gravity describes a universe that behave as an ordinary macroscopic
system. In other words, in both these perspectives, the cosmic evolution could be explained as a tendency for maximizing entropy. Our approach gives a
thermodynamic basis and thus provides further support to the emergent gravity paradigm.

\noindent{\bf Acknowledgements}

We are grateful to the referee for the valuable suggestions which helped for the substantial improvement the manuscript. We are thankful to Prof. M. Sabir for the careful reading
of the manuscript. Thanks are also to IUCAA, Pune for the hospitality during the visit. P. B. Krishna acknowledges KSCSTE, Govt. of Kerala for the financial
support.

%

\end{document}